# On the Performance of Hybrid Search Strategies for Systematic Literature Reviews in Software Engineering

Erica Mourão[a], João Felipe Pimentel[a], Leonardo Murta[a],
Marcos Kalinowski[*,b], Emilia Mendes[c], Claes Wohlin[c]

[a] *Fluminense Federal University, Niterói, RJ, Brazil*
[b] *Pontifical Catholic University of Rio de Janeiro (PUC-Rio), Rio de Janeiro, RJ, Brazil*
[c] *Blekinge Institute of Technology, Karlskrona, Sweden*

**Abstract**

Context: When conducting a Systematic Literature Review (SLR), researchers usually face the challenge of designing a search strategy that appropriately balances result quality and review effort. Using digital library (or database) searches or snowballing alone may not be enough to achieve high-quality results. On the other hand, using both digital library searches and snowballing together may increase the overall review effort.

Objective: The goal of this research is to propose and evaluate hybrid search strategies that selectively combine database searches with snowballing.

Method: We propose four hybrid search strategies combining database searches in digital libraries with iterative, parallel, or sequential backward and forward snowballing. We simulated the strategies over three existing SLRs in SE that adopted both database searches and snowballing. We compared the outcome of digital library searches, snowballing, and hybrid strategies using precision, recall, and F-measure to investigate the performance of each strategy.

Results: Our results show that, for the analyzed SLRs, combining database searches from the Scopus digital library with parallel or sequential snowballing achieved the most appropriate balance of precision and recall.

Conclusion: We put forward that, depending on the goals of the SLR and the available resources, using a hybrid search strategy involving a representative digital library and parallel or sequential snowballing tends to represent an appropriate alternative to be used when searching for evidence in SLRs.

*Keywords:* Systematic literature review; Search Strategy; Database Search; Snowballing; Software Engineering.

## 1. Introduction

Systematic Literature Reviews (SLR) aim at identifying, evaluating, and interpreting relevant research in a specific topic area [1]. Kitchenham and Charters [1], and Wohlin [2] provide guidelines for searching for evidence when conducting an SLR in the SE domain.

The guidelines provided by Kitchenham and Charters [1] recommend composing a string to find relevant studies by searching on several digital libraries (which actually store papers) or index databases (which contain references to papers stored in external digital libraries) to find relevant studies. This SLR search strategy is known as a database search. In such a strategy, both digital libraries and index databases are treated as data sources for finding papers that match a given search string. Therefore, from now on, we consider both types of searches similar and use the terms digital libraries and databases, interchangeably. On the other hand, Wohlin [2] recommends iteratively identifying papers based on the reference list and the citations of papers that were found by an informal database search. This alternative SLR search strategy is known as Backward Snowballing (BS) (when searching within the reference lists) and Forward Snowballing (FS) (when searching within the citing papers).

The database search strategy is the most common and the first published recommendation for SLRs in the SE domain [1]. However, it imposes several challenges for researchers. Some of them include the selection of appropriate digital libraries and the design of a specific search string for conducting searches within those libraries [2]. Furthermore, this strategy needs customization of the search string to allow using it in different digital libraries. Reported difficulties regarding digital libraries include the diversity of user interfaces, the limitation of operators, and not handling synonyms of terms [2]. Moreover, usually, this strategy leads to an overlap of papers in different digital libraries, difficulties concerning the concatenations of keywords, and search execution inconsistencies within specific digital libraries [3].

Snowballing emerged as an attractive alternative or complement to database search. Snowballing does not require searching in more than one digital library; the approach is more understandable and easy to follow [4]. In particular, snowballing is expected to be more efficient when relevant database search keywords include general terms, by reducing the amount of noise. However, it also has drawbacks, such as dependency on an appropriate seed (start) set of relevant papers [2]. Badampudi et al. [5] discuss organizing papers in the seed set into different categories so that the seed set should have at least one paper in each category.

---
[*] Corresponding author. Tel.: +55(21)3527-1510 Ext: 4337; e-mail: kalinowski@inf.puc-rio.br

Furthermore, Jalali and Wohlin [4] discuss difficulties of judgments based on the title of the paper, when applying backward and forward snowballing, which might result in missing papers with no relevant keyword in the title. Additionally, they report that a threat in snowballing is finding several papers from de same authors. Hence, the results might be biased by overrepresenting research from specific authors [4].

Another alternative consists of completely applying database searches in several digital libraries, followed by iterative BS and FS. This alternative improves the chances of identifying relevant studies at the price of adding more effort to the review process. Moreover, it also retains some of the drawbacks of both database searches and snowballing.

Some previous work investigated how to balance evidence quality and effort when conducting SLRs. The studies reported in [5] and [4] compared the database and snowballing search strategies. The study reported in [5] concludes that they are comparable in terms of identifying relevant papers. The authors of [4] argue that snowballing does not require searching in more than one database and that database searches require more effort to refine the searches to identify relevant papers and discard irrelevant ones. Wohlin [2] puts forward that different approaches to identifying relevant literature should be used to ensure the best possible coverage. However, the literature investigating search strategies is scarce, and there is a need to evaluate the performance of different strategies.

Therefore, the goal of this paper is to evaluate the performance of both database and snowballing searches, to define hybrid strategies based on the evaluation, and to evaluate such hybrid strategies. The defined hybrid search strategies regard complementing a single database search on an efficient digital library with four different snowballing variations. When conceiving the hybrid strategies, we chose Scopus for the database search because it was the most efficient digital library when answering RQ1 (see Section 6). The snowballing variations that follow the database search are iterative BS and FS, parallel BS and FS, sequential BS and FS, and sequential FS and BS. The definitions of these strategies can be found in Section 3.

We evaluated the hybrid strategies by simulating their execution over three previously conducted SLRs, [6], [7], and [8], which employed database search in several digital libraries followed by an iterative BS and FS. This simulation allowed us to assess the performance in terms of precision, recall, and F-measure of the hybrid strategies, should they have been used in the original SLR. The results of those four hybrid strategies were compared against popular conventional SLR strategies: pure database search, pure snowballing, and completely combining database searches in several digital libraries followed by iterative BS and FS.

We used the simulation results to provide answers to the following Research Questions (RQs):

*RQ1 - What is the performance of the <u>database search strategy</u> in the published SLRs?*

*RQ2 - What is the performance of the <u>snowballing search strategy</u> in the published SLRs?*

*RQ3 - What is the performance of each <u>hybrid search strategy</u> in the published SLRs?*

The results of our study show that hybrid search strategies have better performance and may be an appropriate alternative compared with database search or snowballing alone.

The remainder of this paper is organized as follows. In Section 2, we provide the background on SLR search strategies. In Section 3, the four hybrid search strategies are defined. In Section 4, we detail our research questions. Section 5 describes the SLR corpus and the evaluation research method. In Section 6, we present the results and answer the research questions. In Section 7, the threats to validity and limitations are discussed. In Section 8, we present the related work and compare with our results. Finally, in Section 9, final remarks and future work are presented.

## 2. Background

According to Kitchenham et al. [1], SLRs must be undertaken by strictly following a predefined search strategy. This search strategy should be unbiased and should allow assessing the completeness of the search. Kitchenham et al. [1] argue that the initial searches for primary studies can be conducted by using several digital libraries, but also indicate that other complementary searches should be employed (e.g., manual searches within proceedings and journals).

One of the challenges of the database searches is to identify terms to formulate an appropriate search string. Furthermore, digital libraries are not designed to support SLRs [1], and an improper selection of search keywords or bugs related to features of the digital libraries could lead to missing relevant papers or retrieving irrelevant papers [3]. Regarding manual searches, they are not commonly used in recently published SLRs and are out of the scope of our investigation.

Wohlin [2] provides guidelines for conducting snowballing as an SLR search strategy. The guidelines define, illustrate, and evaluate snowballing by replicating a published SLR that originally used a database search strategy. The snowballing approach has the challenge of identifying an appropriate seed set of papers. They concluded that snowballing represents an alternative search strategy to use when conducting SLRs, instead of searching in several different databases.

To mitigate the risk of missing relevant evidence, several researchers have combined both search strategies, starting with database searches in several digital libraries and then applying BS and FS iteratively on the set of papers that were selected based on the database searches. Examples of such SLRs include [6], [7], and [8]. Nevertheless, while helping to mitigate the risk of missing relevant evidence, the adoption of database searches in several digital libraries followed by iterative BS and FS might result in significant effort, involving analyzing more irrelevant research papers.

Having this in mind, in our previous work [9], we defined and investigated a hybrid search strategy for selecting studies by combining searches in a specific digital library (Scopus) followed by parallel BS and FS, were papers obtained by backward snowballing are not subject to forward snowballing, and vice-versa. The proposed hybrid search strategy [9] comprised the following four activities: *identify research studies using Scopus database search, select primary studies to compose the seed set, and apply backward and forward snowballing, in parallel, over the seed set*.

We assessed the performance of this hybrid search strategy in terms of precision and recall over two previously conducted SLRs, [10] and [11], which originally employed database searches. Our findings indicated that the proposed hybrid strategy was suitable for the investigated SLRs, providing similar results to using database searches on several digital libraries.

Nevertheless, this preliminary investigation had some significant limitations. First, we only compared the hybrid strategy against the database search strategy, missing comparisons against snowballing and against the exhaustive combined database and snowballing searches. Second, we employed a specific snowballing strategy in which BS and FS are conducted in parallel over the seed set, i.e., the papers obtained by BS are not subject to FS, and vice-versa. This snowballing strategy was introduced as a tentative solution to increase precision without compromising recall [9].

Therefore, in this paper we take the investigation further on hybrid search strategies, more precisely defining different hybrid search strategy possibilities (involving iterative, parallel, and sequential snowballing – *cf.* Section 3) and evaluating them against database searches, snowballing searches, and exhaustive combined database and snowballing searches. As a basis for comparisons, the evaluations were performed on three different SLRs, [6], [7] and [8], which employed complete combined database and iterative snowballing searches.

## 3. Hybrid Search Strategies

Herein we propose four hybrid search strategies and contrast them with three baseline strategies (database search [1], snowballing [2], and database search followed by exhaustive iterative BS and FS). These hybrid strategies combine database search over one specific digital library with different variations of the snowballing steps. As previously stated, the adopted database search for the hybrid strategies was Scopus, because it was the most efficient digital library when answering RQ1 (see Section 6). Moreover, the snowballing steps comprise running backward and forward snowballing in iterations (the strategy defined in the original snowballing guidelines by Wohlin [2]), in parallel (where papers obtained by BS are not subject to FS, and vice-versa), and sequentially (applying the FS iterations after finishing all BS iterations, or vice-versa). Table 1 lists all seven strategies analyzed in this paper. The first three represent the baseline strategies used for comparisons, while the remaining four represent the proposed hybrid strategies.

Table 1 - Analyzed search strategies.

| Strategy | Hybrid Strategy | Description |
|---|---|---|
| DB Search | | Database search (set of digital libraries) |
| SB Search (BS*FS) | | Snowballing search = Google Scholar + iterative backward & forward snowballing |
| DB Search + BS*FS | | Database search (set of digital libraries) + iterative backward & forward snowballing |
| Scopus + BS*FS | x | Scopus database search + iterative backward & forward snowballing |
| Scopus + BS\|\|FS | x | Scopus database search + parallel backward & forward snowballing |
| Scopus + BS+FS | x | Scopus database search + sequential backward & forward snowballing |
| Scopus + FS+BS | x | Scopus database search + sequential forward & backward snowballing |

**DB Search**: this strategy follows the usual database search guidelines [1]. The selected papers come from different queries over multiple digital libraries.

**SB Search (BS*FS)**: this strategy follows the usual snowballing guidelines [2]. It starts with an informal search in Google Scholar to compose a seed set. Then, multiple iterations of backward and forward snowballing are iteratively applied to the seed set to find other papers.

**DB Search + BS*FS**: this strategy combines both the search over all digital libraries and the full-fledged iterative snowballing, respectively described in the *DB Search* and *SB Search (BS*FS)* strategies. First, we perform searches over different digital libraries to compose our seed set. Then, we apply iterative backward and forward snowballing over the seed set and the results obtained by the snowballing.

**Scopus + BS*FS**: this strategy first runs a search over Scopus to compose a seed set. Then, other papers are obtained from the seed set via iterative backward and forward snowballing.

**Scopus + BS\|\|FS**: this strategy also starts with a search over Scopus to compose a seed set. Then, backward and forward snowballing run in parallel over the same seed set. In other words, the papers obtained by backward snowballing are not subject to forward snowballing, and vice-versa. This strategy was first introduced by Mourão et al. [9] as an attempt to increase precision without compromising recall.

**Scopus + BS+FS**: similar to the previous strategies, this strategy also starts with a search over Scopus to compose a seed set. Then, multiple iterations of backward snowballing occur over the seed set. After finishing all backward snowballing iterations, forward snowballing starts. In this strategy, the papers obtained by forward snowballing are not subject to backward snowballing.

**Scopus + FS+BS**: this strategy also starts with a search over Scopus to compose a seed set. Then, multiple iterations of forward snowballing occur over the seed set. After finishing all forward snowballing iterations, backward snowballing starts. In this strategy, the papers obtained by backward snowballing are not subject to forward snowballing.

In this paper, we simulated each strategy in the supporting tool. For the snowballing steps, BS navigates in the references list of a given paper and, FS uses Google Scholar to obtain the papers that cite a given paper.

## 4. Research Questions

Our first research question – *RQ1 - What is the performance of the <u>database search strategy</u> in the published SLRs?* – focuses on analyzing the performance of the database search component alone in the published SLRs. We split this question into three sub-questions:

*RQ1.1 - What is the performance of each digital library in the published SLRs?* Digital libraries are different; some are more selective, returning only papers that are very adherent to the query. Others are more inclusive, returning many papers that may or may not fit the researchers' needs. In this RQ, we contrast the performance of digital libraries in terms of precision, recall, and F-measure. The precision is the percentage of papers retrieved by the digital library that was selected by the SLR. The recall is the percentage of selected papers of the SLR that were retrieved by the digital library. The F-measure is the harmonic mean between precision and recall, indicating a compromise between precision and recall.

*RQ1.2 - How many papers of the published SLRs are indexed in the digital libraries?* The search interface, the search engine, and the search string are not perfect. Sometimes, papers that are indexed in the digital library are not retrieved. In this RQ, we do a direct search for the title of each paper in the digital library to evaluate their recall regardless of the search string. This measure indicates the percentage of the papers indexed by the digital library that could have been retrieved by the SLR.

*RQ1.3 - How complementary or overlapping are the digital libraries in the published SLRs?* The search in several digital libraries consumes a considerable amount of effort. Considering that some papers are indexed in more than one digital library, this overlap could represent unnecessary effort for researchers. In this RQ, we evaluate whether some digital libraries used in the published SLR are redundant, retrieving the same set of papers as other digital libraries.

Our second research question – *RQ2 - What is the performance of the <u>snowballing search strategy</u> in the published SLRs?* – focuses on analyzing the performance of the snowballing component alone in the published SLRs. It was also split into three sub-questions:

*RQ2.1 - What is the performance of complementing digital library searches with snowballing in the published SLRs?* Automated search in digital libraries is the most common strategy for conducting SLRs. However, difficulties in creating appropriate search strings and the quality of the search engines may jeopardize the SLR. Thus, snowballing over the reference list and citations of selected papers may help to identify other relevant studies and complement the digital library search. In this RQ, we measure the precision, recall, and F-measure of each forward and backward snowballing iteration.

*RQ2.2 - How complementary or overlapping are backward and forward snowballing in the published SLRs?* Both backward and forward snowballing are important for finding relevant papers. However, there is a risk to have overlap among the set of papers retrieved by backward and forward snowballing, due to the publication dates of different papers identified. On the other hand, given the seed set, one could miss relevant papers by choosing to do just backward or forward snowballing alone. To identify if backward and forward snowballing is complementary or overlapping, we first simulate forward and backward snowballing independently, to collect the set of papers that could have been obtained by each one of them. Then, we investigate the intersection between these sets.

Our third research question – *RQ3 - What is the performance of each <u>hybrid search strategy</u> in the published SLRs?* – focuses on analyzing the performance of the proposed hybrid search strategies in the published SLR. A hybrid strategy combines specific variations of database search and snowballing, eventually focusing on result quality in detriment of review effort, or vice versa. For this RQ, we measure the precision, recall and F-measure of each hybrid strategy, and contrast to the values obtained with the baseline strategies (digital library search, snowballing, and the exhaustive combination of both).

## 5. Search Strategy Evaluation

In this section, we provide the details of how we conducted our evaluation. However, first we describe how we selected the SLRs that were used to simulate the different strategies, the evaluation procedure, and the supporting tool implemented to automate the simulations.

### 5.1. SLR Selection

To answer the research questions, we needed high-quality SLRs that had all the necessary information for the intended simulations. We first performed a search for SLRs that mention both database search and snowballing. We identified a set of seven candidate SLRs. We then evaluated the quality of the SLRs using the same quality criteria used by Kitchenham et al. [12] in their tertiary study on SLRs:
- Are the review's inclusion and exclusion criteria described and appropriate (QA1)?
- Is the literature search likely to have covered all relevant studies (QA2)?
- Did the reviewers assess the quality/validity of the included studies (QA3)?
- Were the basic data/studies adequately described (QA4)?

Additionally, we had to formulate one more criterion to ensure that the selected studies had all the necessary information for simulating the hybrid strategies:
- Does the SLR combine database search with iterative backward and forward snowballing (QA5)?

We score the questions, as suggested by Kitchenham [13]:

*QA1*: <u>Yes</u>, the inclusion criteria are explicit; <u>Partly</u>, the inclusion criteria are implicit; and <u>No</u>, the inclusion criteria are not defined.

*QA2*: <u>Yes</u>, the authors have either searched four or more digital libraries and included additional search strategies or identified and referenced all journals addressing the topic of interest; <u>Partly</u>, the authors have searched three or four digital libraries with no extra search strategies or searched a defined but restricted set of journals and conference proceedings; and <u>No</u>, the authors have searched up to two digital libraries or an extremely restricted set of journals.

*QA3*: <u>Yes</u>, the quality criteria are explicit, and they were applied to each primary study; <u>Partly</u>, the research question involves quality issues that are addressed by the study; and <u>No</u>, the quality criteria are not defined.

*QA4*: <u>Yes</u>, information about each study is described; <u>Partly</u>, only summary information about the set of studies is described; and <u>No</u>, the basic information about the studies were not described.

*QA5*: <u>Yes</u>, the authors applied database search and complemented it with iterative backward and forward snowballing; <u>Partly</u>, the authors applied a database search and complemented it with either backward or forward snowballing; and <u>No</u>, the authors applied database search and snowballing, but snowballing was not directly applied to the results of the database search.

The scoring procedure was Yes = 1, Partly = 0.5, and No = 0. Table 2 shows the score for each of the seven candidate SLRs. The results of the quality assessment show that, while all studies scored 1 in questions QA1 to QA4, only three studies scored 1 in our additional question QA5.

Studies S4 [14] and S5 [15] applied the database search strategy and manual searches, then, after merging the results of these strategies, applied backward snowballing alone. Study S6 [16] was conducted using a database search and complemented with backward snowballing alone. Study S7 [17] used database search and manual searches and, after merging the results of these strategies, complemented them with backward and forward snowballing.

Table 2 - Quality evaluation of the candidate SLRs.

| Study | Ref | QA1 | QA2 | QA3 | QA4 | QA5 | Score |
|-------|-----|-----|-----|-----|-----|-----|-------|
| S1 | [6] | 1 | 1 | 1 | 1 | 1 | 5 |
| S2 | [7] | 1 | 1 | 1 | 1 | 1 | 5 |
| S3 | [8] | 1 | 1 | 1 | 1 | 1 | 5 |
| S4 | [14] | 1 | 1 | 1 | 1 | 0 | 4 |
| S5 | [15] | 1 | 1 | 1 | 1 | 0 | 4 |
| S6 | [16] | 1 | 1 | 1 | 1 | 0.5 | 4.5 |
| S7 | [17] | 1 | 1 | 1 | 1 | 0 | 4 |

Studies S1 [6], S2 [7], and S3 [8] are SLRs that are completely compliant to QA5, and hence, they were selected to compose the corpus of our study. They combine database search in several digital libraries with iterative backward and forward snowballing. We detail each of the selected SLRs, hereafter referred to as SLR1, SLR2, and SLR3, in the following.

SLR1 [6] investigates evidence on approaches for the strategic alignment of software process improvement (SPI). It started with a digital library search on Springer, Scopus, Web of Science, Science Direct, Compendex, IEEE Xplore, and ACM Digital Library, followed by iterative backward and forward snowballing. It was published in 2017, with a database search conducted in August 2015, without limiting the publication year. Snowballing was performed in July 2016. SLR1 selected 51 studies in total, where 22 came from the database search and 29 from snowballing. The study has an additional quality assessment step after snowballing, resulting in a final dataset with 30 papers. However, in terms of the traditional SLR process [1], the search strategy concerns the study identification, and we used the study selection step (i.e., the application of the inclusion and exclusion criteria) as the basis for our assessment of precision and recall. Indeed, the search strategy has limited influence on the quality of the studies, and the quality assessment is a separate and later step of the SLR process [1], typically influenced by specific SLR goals. We decided using the 51 papers obtained before this step of quality assessment for the evaluations of the search strategies. This decision is important because all 51 papers were subject to snowballing, having a direct effect on the effort of the SLR. Not considering all of them would jeopardize the precision, recall, and F-metric measures. The list of selected papers of SLR1 is available on our companion website[1].

SLR2 [7] aims at identifying and making a synthesis of the Definition of Done (DoD) criteria used in agile software development projects. It was published in 2017, with the database search conducted in June 2016 and the snowballing conducted in August 2016. The search strategy involved database searches on ACM Digital Library, Engineering Village (Compendex), Science Direct, Scopus, Springer, Web of Science, and Wiley Online Library. After that, the authors performed snowballing on the set of papers selected as a result of the database search. SLR2 selected 20 research papers, where 16 came from the database search and four from snowballing. After that, the authors applied an additional quality assessment and ended up with eight papers. Following the same argumentation as for SLR1, we decided to use the 20 papers obtained before this step of quality assessment for the evaluation of the search strategies. The list of selected papers of SLR2 is also available on our companion website[1].

SLR3 [8] investigates the use and usefulness of ontologies in software process assessment (SPA). It was published in 2017, with database search and snowballing conducted in December 2016 and January 2017, respectively. The search strategy involved database searches on ACM Digital Library, Google Scholar, IEEE Xplore, Science Direct, Scopus, Springer, Web of Science, and Wiley Online Library. Afterward, the authors performed snowballing on the set of papers selected by the database search. SLR3 selected 14 research papers, including eleven papers from the database search and three papers from snowballing. The list of selected papers of SLR3 is also available on our companion website[1].

Table 3 - Digital Libraries used for each SLR.

| Digital Library | SLR1 | SLR2 | SLR3 |
|-----------------|------|------|------|
| ACM Digital Library | x | x | x |
| Compendex | x | x | - |
| Google Scholar | - | - | x |
| IEEE Xplore | x | - | x |
| Science Direct | x | x | x |
| Scopus | x | x | x |
| Springer | x | x | x |
| Web of Science | x | x | x |
| Wiley Online Library | - | x | x |

We analyzed the digital libraries adopted by the SLRs that compose our corpus. Table 3 shows the use of the digital library in an SLR, denoted by "x", and the absence, denoted by "-". It is possible to observe that, while each SLR used a different set of digital libraries, they used at least seven different digital libraries each and have several in common.

---

[1] https://gems-uff.github.io/hybrid-strategies

Finally, as a summary of the study selection in the three SLRs, Table 4 shows that together they returned 2,803 papers in the database search. However, 891 were duplicates, leading to a total of 1,912 unique papers. Out of those 1,912 unique papers, 49 were selected from the database search and 36 were selected from snowballing.

Table 4 - Study selection summary.

| SLR | Papers with duplicates | Papers without duplicates | Seed set | Snowballing |
|---|---|---|---|---|
| SLR1 | 517 | 497 | 22 | 29 |
| SLR2 | 1,715 | 935 | 16 | 4 |
| SLR3 | 571 | 480 | 11 | 3 |

### 5.2. Evaluation Procedure

After selecting the SLRs that compose our corpus, our evaluation procedure comprises three main activities: data extraction, strategies simulation, and analysis. Hereafter we describe each of these activities.

#### 5.2.1. Data Extraction

For each paper, we first extracted the data provided in the paper (e.g., the list of digital libraries, the number of visited papers in each digital library, and the number of selected papers). Then, we contacted the authors requesting additional data. Finally, if the data was not complete enough to reproduce the results, we rerun the queries over the digital libraries.

The authors provided us spreadsheets with the studies listed in the published SLR. From each published SLR, we extracted the number of visited papers in each digital library, the number of primary studies selected in the SLR, the search string used, the list of digital libraries used, and the number of duplicate papers. For each paper, we identified the title, year of publication, authors, type of paper, publisher of the selected studies, the references list, and the paper's citations. Moreover, we tagged the papers, identifying them as seed set (prevenient from database search) or as snowballing, and if it was selected in the SLR.

In some cases, the authors did not document duplicates. For example, a paper that was obtained from both ACM Digital Library and Scopus should have been registered as both ACM Digital Library and Scopus and marked as a duplicate. However, some authors just stated the first digital library that returned the paper. This could jeopardize our results, as we needed to count the visited and selected papers for each digital library to calculate the precision and recall. In these cases, we rerun the search in each digital library to check whether there were missing duplicates.

#### 5.2.2. Strategies Simulation

We simulate the hybrid search strategies using a snowballing supporting tool that we created (*cf.* Section 5.3) and applying it to the three published SLR of our corpus. The following steps compose the process applied to each SLR.

**1. Extract data from each paper.** A total of 2,803 papers is identified based on the database search in all SLRs (*cf.* Table 4). We extracted the data of interest from each paper to insert it into the supporting tool.

**2. Insert data into the supporting tool.** We inserted the data of the paper and a tag with its provenance in the supporting tool. During this process, the tool identified and removed 891 duplicates. Nevertheless, the tool registers the set of digital libraries in which each paper was found and information about whether the paper was in the seed set of papers or retrieved through snowballing.

**3. Simulate search strategies.** We simulate each of the baseline and hybrid search strategies in the supporting tool. We simulated the strategies independently: *DB Search; SB Search; DB Search + BS*FS; Scopus + BS*FS; Scopus + BS//FS; Scopus + BS+FS; and Scopus + FS+BS*. We focused on reproducing equivalent results of the original SLRs, limited to identifying the papers that were selected in the original SLRs within the seed set and in the iterations of snowballing.

**4. Calculate metrics.** Finally, the tool calculates the metrics precision, recall, and F-measure based on the number of visited and selected papers for each search strategy. We compared these results to analyze the performance of each search strategy.

#### 5.2.3. Analysis (Answering the RQs)

Aiming at answering our RQs, we used the strategy as an independent variable and precision, recall, and F-measure as dependent variables. Precision and recall are standard information retrieval metrics [18] used to compare results with a predefined oracle. Originally, precision indicates the fraction of retrieved documents that are known to be relevant, recall indicates the fraction of known relevant documents that were effectively retrieved, and the F-measure indicates the harmonic mean between precision and recall.

In our context, the oracle is the set of selected papers from the SLRs (strategy *DB Search + BS*FS*). Precision indicates the correctness of a given strategy in finding appropriate papers. For example, a 100% precision would indicate that all papers visited by the strategy were selected by the SLR. Recall indicates the completeness of a given strategy. For example, a 100% recall would indicate that all selected papers of the SLR were visited by the strategy. Hence, these metrics balance the two main aspects in SLR search strategies: precision represents results' correctness and recall represents results' completeness. Finally, the F-measure indicates the best compromise between precision and recall. Formally, the metrics are defined as follows:

$$Precision = \frac{Visited \cap Selected}{Visited}$$

$$Recall = \frac{Visited \cap Selected}{Selected}$$

$$F\text{-}measure = 2 \times \frac{Precision \times Recall}{Precision + Recall}$$

*5.3. Supporting Tool*

Simulating a systematic literature review is not trivial. Another researcher using the SLR protocol should be able to reproduce the same results. It requires keeping track of the visited papers and avoiding duplicates of them for avoiding rework. Simulating each strategy to compare them manually is even harder, as it requires managing the same papers multiple times.

To reduce this effort and potential errors, we developed a snowballing supporting tool[2] composed of a set of Jupyter notebooks and Python scripts to manage SLR papers and allow simulating the strategies. These scripts enable researchers to register papers without duplicates, find forward citations, and analyze the results of the snowballing. Moreover, after registering all papers obtained from the exhaustive database search followed by iterative snowballing, the scripts allow the simulation of all other strategies.

The usage of the scripts can be divided into two phases: registration and analysis. In the first phase, the researcher registers papers and then the tool provides support for performing both the backward and forward snowballing. In the second phase, the researcher analyzes the results. Since the scripts can collect the citations, they can generate citation graphs for the simulated strategies.

In the first phase, researchers can use Jupyter notebooks to register papers from the seed set and from backward snowballing into the tool. In our case, we had the snowballing results of the published SLRs. Thus, we inserted all studies included in the original SLR. In addition to these papers, we also performed backward snowballing on the selected papers and inserted the results. For the backward snowballing, we extracted the references to BibTeX and used the notebooks to insert the papers together with their citations.

The next step consists of applying forward snowballing over the selected papers and registering studies that cite those papers. Note that the scripts use only Google Scholar to find the citations during forward snowballing. We chose Google Scholar because it has a consistent collection of papers that cite a given paper and for being effective and widely adopted for forward snowballing purposes [22]. We repeat all steps of the process for each selected study of each SLR.

In addition to registering the papers, we also included tags in each paper to indicate their provenance, i.e., which digital libraries return them, which papers belong to the seed set, which papers were selected by the SLRs, and which papers were obtained through snowballing. We used these tags afterward to support the analyses.

In the second phase, researchers can use Jupyter notebooks to analyze papers' metadata, generate provenance and citation graphs, and describe the snowballing process through simulations of strategies' iterations based on citations. Thus, we automated the calculation of the measures for each hybrid search strategy. All Jupyter notebooks are available on our companion website[1].

# 6. Results and Discussion

In this section, we answer the research questions defined in Section 4.

*6.1. What is the performance of each digital library in the published SLRs (RQ1.1)?*

Our results show that the database search strategy visited 497 papers retrieved from digital libraries in SLR1, 935 papers in SLR2, and 480 in SLR3. Moreover, among the visited papers, we identified 51 studies selected by SLR1, 20 studies selected by SLR2, and 14 selected by SLR3. After that, as shown in Table 5, we calculate the performance of each digital library in terms of precision, recall, and F-measure.

Concerning precision, we can observe that Compendex provides the highest value for SLR2 and the third-highest value for SLR1 (SLR3 did not use Compendex) when compared to the other digital libraries. Although Web of

Table 5 - Performance of digital libraries in the SLRs.

| Digital Library | Precision (%) | | | Recall (%) | | | F-measure (%) | | |
|---|---|---|---|---|---|---|---|---|---|
| | SLR1 | SLR2 | SLR3 | SLR1 | SLR2 | SLR3 | SLR1 | SLR2 | SLR3 |
| ACM Digital Library | 5.00 (5/100) | 2.38 (5/210) | NAN (0/0) | 9.80 (5/51) | 25.00 (5/20) | 0.00 (0/14) | 6.62 | 4.35 | 0.00 |
| Compendex | 38.46 (5/13) | 25.00 (2/8) | - | 9.80 (5/51) | 10.00 (2/20) | - | 15.62 | 14.29 | - |
| Google Scholar | - | - | 2.36 (11/466) | - | - | 78.57 (11/14) | - | - | 4.58 |
| IEEE Xplore | 13.95 (6/43) | - | NAN (0/0) | 11.76 (6/51) | - | 0.00 (0/14) | 12.77 | - | 0.00 |
| ScienceDirect | 0.51 (1/195) | 2.01 (5/249) | 0.00 (0/21) | 1.96 (1/51) | 25.00 (5/20) | 0.00 (0/14) | 0.81 | 3.72 | 0.00 |
| Scopus | 46.67 (7/15) | 9.09 (7/77) | 3.80 (3/79) | 13.73 (7/51) | 35.00 (7/20) | 21.43 (3/14) | 21.21 | 14.43 | 6.45 |
| Springer | 1.42 (2/141) | 0.81 (1/124) | 1.54 (1/65) | 3.92 (2/51) | 5.00 (1/20) | 7.14 (1/14) | 2.08 | 1.39 | 2.53 |
| Web of Science | 50.00 (5/10) | 0.00 (0/3) | NAN (0/0) | 9.80 (5/51) | 0.00 (0/20) | 0.00 (0/14) | 16.39 | 0.00 | 0.00 |
| Wiley Online Library | - | 0.00 (0/295) | 0.00 (0/15) | - | 0.00 (0/20) | 0.00 (0/14) | - | 0.00 | 0.00 |

---

[2] https://github.com/JoaoFelipe/snowballing

Science showed the highest result for SLR1, it had very poor results for SLR2 and SLR3. Finally, Scopus consistently delivered high values for precision: the second-highest result for SLR1 and SLR2 and the highest result for SLR3.

Regarding recall, Scopus is again a prominent option – it provides the highest results for SLR1 and SLR2, and the second-highest for SLR3. No other digital library was consistently efficient regarding recall in all three SLRs. Google Scholar showed the highest recall for SLR3 (SLR1 and SLR2 did not use Google Scholar). This result is not surprising, considering that Google Scholar is not exactly a digital library, but a search engine that references multiple digital libraries.

Finally, regarding the F-measure, Scopus was undoubtedly the most prominent digital library, with the highest values for all three SLRs. Compendex also appears as a strong competitor, with the third-highest result for SLR1 and second-highest for SLR2 (almost tied with Scopus). Google Scholar also appears with the second-highest value for SLR3. No other digital library consistently provided high values in terms of the F-measure.

This observed result of precision and recall is expected, considering that Scopus, Compendex, Google Scholar, and Web of Science are in fact index databases. In other words, they store references to papers stored in external digital libraries, rather than the papers per se. Thus, they cover a wide range of publishers and are naturally able to reach more papers than digital libraries that belong to a specific publisher.

**Answer to RQ1.1:** Scopus and Compendex (only SLR1 and SLR2) are prominent options in terms of precision. Scopus and Google Scholar (only SLR3) are prominent options in terms of recall. When considering precision and recall together (F-measure), Scopus and Compendex stood out, with Scopus clearly ahead.

**Implications:** Scopus is a consistent option, but it finds just a limited amount (13% to 35%) of relevant papers alone. Thus, complementing Scopus with other digital libraries or snowballing is necessary.

*6.2. How many papers of the published SLRs are indexed in the digital libraries (RQ1.2)?*

After querying using the title of each selected paper from the SLRs on every digital library, we could observe in Table 6 that Compendex delivered the highest recall result for SLR2 (a tie with Scopus) and the second-highest for SLR1 (SLR3 did not use Compendex). Google Scholar provides the highest value for SLR3 (SLR1 and SLR2 did not use Google Scholar). Finally, Scopus delivered again the highest result for SLR1, SLR2 (tied with Compendex), and second-highest result for SLR3 it still seems to be an interesting trade-off, in particular considering that high recall is important in an SLR. It is noteworthy that stopping the SLRs in the database search phase would have retrieved just 43% to 80% of the papers. One snowballing iteration increases this recall to range between 90.2% to 100% of the papers.

Most of the digital libraries had a substantial increase from the concrete recall shown in Table 5 to the potential recall shown in Table 6. However, Wiley Online Library did not provide any increase in results for SLR2 and SLR3 (SLR1 did not use Wiley Online). ScienceDirect provided a small increase for all SLRs. Moreover, ACM Digital Library showed a very subtle increase for SLR1 and did not show any increase for SLR2 and SLR3.

Again, Scopus, Compendex, Google Scholar, and Web of Science, which are index databases, achieved the highest potential recall in our analysis. As previously explained, index databases are not restricted to a specific publisher, being able to return papers that reside on different external digital libraries.

Table 6 - Recall of papers indexed in the published SLR.

| Digital Library | Recall of indexed (%) | | |
|---|---|---|---|
| | SLR1 | SLR2 | SLR3 |
| ACM Digital Library | 11.76 (6/51) | 25.00 (5/20) | 0.00 (0/14) |
| Compendex | 68.63 (35/51) | 95.00 (19/20) | - |
| Google Scholar | - | - | 100.00 (14/14) |
| IEEE Xplore | 27.45 (14/51) | - | 21.43 (3/14) |
| ScienceDirect | 3.92 (2/51) | 30.00 (6/20) | 7.14 (1/14) |
| Scopus | 82.35 (42/51) | 95.00 (19/20) | 50.00 (7/14) |
| Springer | 31.37 (16/51) | 10.00 (2/20) | 21.43 (3/14) |
| Web of Science | 52.94 (27/51) | 55.00 (11/20) | 35.71 (5/14) |
| Wiley Online Library | - | 0.00 (0/20) | 0.00 (0/14) |

**Answer to RQ1.2:** Scopus and Compendex are prominent options in terms of potential recall, with Scopus ahead. A side note is for Google Scholar that reached 100% recall in the only SLR that adopted it (SLR3).

**Implications:** Scopus could have found alone 50% to 95% of the papers. This motivates extra effort on the elaboration of search strings, considering the gap between the potential recall and the concrete recall delivered by the SLRs. Finally, as the sum of potential recall for each SLR surpasses 100%, a wise choice of the digital libraries is necessary to avoid rework (many duplicates)

*6.3. How complementary or overlapping are the digital libraries in the published SLRs (RQ1.3)?*

To answer this question, as shown in Table 7, we compared the set of selected papers among each pair of digital libraries. For example, all ten papers selected by ACM Digital Library are unique. Consequently, ACM Digital Library complements other digital libraries for SLR1 and SLR2 (no paper was identified for SLR3). On the other hand, six papers out of seven selected by Compendex are duplicate. Consequently, Compendex was mostly redundant.

We highlight in bold all cases with more than 50%. Google Scholar was able to retrieve all papers retrieved by Scopus and Springer. However, as previously mentioned, just SLR3 used Google Scholar and more evidence is needed. Moreover, Scopus was able to retrieve most of the papers retrieved by Compendex (85%) and Web of Science (80%). Finally, Compendex was able to retrieve 60% of the papers retrieved by Web of Science.

These results can be better understood when considering that Scopus and Compendex are both index databases and belong to the same publisher: Elsevier. Moreover, Google Scholar and Web of Science are also index databases. Although they belong to different companies, all four display the highest overlaps in our study, as expected.

As shown in Table 7, NAN means Not a Number. For example, 0.0 (zero) divided by 0.0 (zero) is arithmetically undefined, and it is denoted as NAN. As expected, the intersection between the digital library and index database only occurs when the index database is involved.

**Answer to RQ1.3:** ACM Digital Library is complementary to other digital libraries, and Scopus contains most of the results provided by Compendex and Web of Science. Google Scholar contained both Scopus and Springer, but the evidence came from just one SLR.

**Implications:** When Scopus is adopted, also adopting Compendex and Web of Science does not seem to be highly relevant. On the other hand, the ACM Digital Library should be considered. However, Scopus and ACM Digital Library were not able to reach all papers. Thus, complementing digital library search with snowballing is recommended.

*6.4. What is the performance of complementing digital library searches with snowballing in the published SLRs (RQ2.1)?*

As shown in Table 8, complementing the Scopus search with one iteration of snowballing was able to provide a 100% recall for SLR2 and SLR3, without a huge loss of precision (reductions from 1.71% to 1.29% for SLR1 and from 2.29% to 1.81% for SLR2). For SLR1, one snowballing iteration increases the recall from 43.14% to 90.2%, while precision decreases from 4.73% to 3.72%. Hence, while the F-measure slightly decreases with one snowballing iteration, it still seems to be an interesting trade-off, in particular considering that high recall is important in an SLR. It is noteworthy that stopping the SLRs in the database search phase would have retrieved just 43% to 80% of the papers. One snowballing iteration increases this recall to range between 90.2% to 100% of the papers.

Figure 1, Figure 2, and Figure 3 complement Table 8 by showing a visual representation (generated by our supporting tool) of the whole search process for SLR1, SLR2, and SLR3, respectively. They capture the number of visited and selected papers for each of the digital libraries and the snowballing iterations. These visualizations reinforce the need of at least one snowballing iteration, in particular when a limited number of digital libraries is used.

**Answer to RQ2.1:** A single iteration of backward and forward snowballing complementing the database search, helped to obtain 100% of recall for SLR2 and SLR3, and 90% recall for SLR1.

**Implications:** Although the first iteration of snowballing is positive, we could only observe a 100% recall for all three SLR after three iterations of snowballing, with negative consequences for precision. Thus, analyzing the interplay of backward and forward snowballing may help in devising more efficient hybrid strategies.

*6.5. How complementary or overlapping are backward and forward snowballing in the published SLRs (RQ2.2)?*

As shown in Table 8, regarding precision, we can observe that forward snowballing provides the highest value in the iterations concerning the total number of selected papers. However, the difference between forward and backward is low. Regarding recall, in the first iteration, backward snowballing provides the highest value for all SLRs. Forward snowballing provides the highest value in the following iterations for SLR1, after SLR2 and SLR3 reached 100%.

Figure 4 visually contrasts the sets of papers retrieved using forward and backward snowballing. For SLR1, the final set of papers includes 17 found via backward and 16 via forward with four identical papers. This means a 24% overlap for backward and 25% for forward snowballing. Thereby, backward and forward snowballing are complementary. For SLR2, the final set of papers includes four papers found via backward, one of them is also found via forward snowballing. For SLR3, the final set includes three papers found via backward, one of them is also found via forward snowballing. In the cases of SLR2 and SLR3, backward includes all papers found via forward snowballing.

Table 7 - Complementing versus overlapping digital libraries in the published SLRs.

| Row contains Column (diagonal indicates unique papers) | ACM Digital Library | Compendex | Google Scholar | IEEE Xplore | Science Direct | Scopus | Springer | Web of Science | Wiley Online Library |
|---|---|---|---|---|---|---|---|---|---|
| **ACM Digital Library** | **10/10** | 0% (0/7) | 0% (0/11) | 0% (0/6) | 0% (0/6) | 0% (0/17) | 0% (0/4) | 0% (0/5) | NAN (0/0) |
| **Compendex** | 0% (0/10) | **1/7** | - | 16% (1/6) | 0% (0/6) | 42% (6/14) | 33% (1/3) | **60% (3/5)** | NAN (0/0) |
| **Google Scholar** | NAN (0/0) | - | **7/11** | NAN (0/0) | NAN (0/0) | **100% (3/3)** | **100% (1/1)** | NAN (0/0) | NAN (0/0) |
| **IEEE Xplore** | 0% (0/5) | 20% (1/5) | 0% (0/11) | **5/6** | 0% (0/1) | 10% (1/10) | 0% (0/3) | 0% (0/5) | NAN (0/0) |
| **ScienceDirect** | 0% (0/10) | 0% (0/7) | 0% (0/11) | 0% (0/6) | **5/6** | 5% (1/17) | 0% (0/4) | 0% (0/5) | NAN (0/0) |
| **Scopus** | 0% (0/10) | **85% (6/7)** | 27% (3/11) | 16% (1/6) | 16% (1/6) | **6/17** | 25% (1/4) | **80% (4/5)** | NAN (0/0) |
| **Springer** | 0% (0/10) | 14% (1/7) | 9% (1/11) | 0% (0/6) | 0% (0/6) | 5% (1/17) | **2/4** | 0% (0/5) | NAN (0/0) |
| **Web of Science** | 0% (0/10) | 42% (3/7) | 0% (0/11) | 0% (0/6) | 0% (0/6) | 23% (4/17) | 0% (0/4) | **1/5** | NAN (0/0) |
| **Wiley Online Library** | 0% (0/5) | 0% (0/2) | 0% (0/11) | NAN (0/0) | 0% (0/5) | 0% (0/10) | 0% (0/2) | NAN (0/0) | **NAN (0/0)** |

**Answer to RQ2.2:** Forward snowballing is a prominent option in terms of precision. Backward snowballing is a prominent option in terms of recall. In SLR1, backward and forward were complementary. In SLR2 and SLR3, the backward snowballing included all papers retrieved by the forward snowballing.

**Implications:** Both backward and forward snowballing contribute to the precision and recall of SLRs. However, one or two iterations may suffice to reach relevant recall values.

*6.6. What is the performance of each hybrid search strategy in the published SLRs (RQ3)?*

As shown in Table 9, regarding precision, we can observe that *Scopus + BS+FS* provides the highest value for SLR2 (tied with *Scopus + BS//FS*), the second-highest value for SLR1, and the third-highest value for SLR3, when compared to the other strategies.

*Scopus + FS+BS* provides the second-highest value for SLR2 and SLR3, and the third highest value for SLR1. However, the strategy *Scopus + BS//FS* consistently provides the highest value for SLR1, SLR2, and SLR3.

On the other hand, regarding the recall, the strategies *Scopus + BS//FS*, *Scopus + FS+BS*, and *Scopus + BS+FS* show only the third or the fourth-highest values. *Scopus + BS*FS* provides the second-highest recall for SLR1, SLR2, and SLR3. Of course, the highest recall is achieved by *DB Search + BS*FS* for all three SLRs, as it refers to the complete strategy, and the one used in the baseline of our corpus to identify the overall selected papers.

Finally, regarding the F-measure, *Scopus + BS//FS* provides the highest value for SLR2 and SLR3, and *Scopus + BS+FS* the highest values for SLR1 and SLR2 (the later one tied with *Scopus + BS//FS*).

**Answer to RQ3:** *Scopus + BS//FS* and *Scopus + BS+FS* strategies are prominent options in terms of precision. *DB Search + BS*FS* and Scopus + BS*FS* are prominent options in terms of recall. When considering precision and recall together (F-measure), *Scopus + BS//FS* and *Scopus + BS+FS* stood out. Only *DB Search + BS*FS* provides 100% of recall, but with low precision, typically requiring significant effort.

**Implications:** *Scopus + BS//FS*, *Scopus + BS+FS*, and *Scopus + FS+BS* demand less effort than plain *DB Search* and *DB Search + BS*FS*, but with a price in recall. Depending on the goals of the SLR and the resources available, one of these hybrid strategies may be an appropriate alternative. However, no other strategy besides *DB Search + BS*FS* could consistently guarantee high levels of recall.

## 7. Threats to Validity and Limitations

Although we aimed at reducing the threats to validity of our study, some decisions may have affected the results. We discuss the threats to validity, based on the types presented in [19] hereafter.

Regarding **construct validity**, we adopted precision, recall, and the F-measure to assess the performance of the search strategies. Precision and recall are commonly used in other studies that investigate search strategies [5, 9, 21, 25, [27]]. Mourão et al. [9], Felizardo et al. [21], and Skoglund and Runeson [25] use precision and recall. Dieste and Padua [26] and Zhang and Ali Babar [27] use precision and sensitivity, with the same meaning as our precision and recall. Finally, Badampudi et al. [5] do not formally define the metrics but use them to calculate the total efficiency (precision) and the percentages of studies identified (recall).

*Table 8* - Performance of complementing digital libraries with snowballing for the analyzed SLRs.

| Iteration | State | Accumulated Precision (%) | | | Accumulated Recall (%) | | | Accumulated F-measure (%) | | |
|---|---|---|---|---|---|---|---|---|---|---|
| | | SLR1 | SLR2 | SLR3 | SLR1 | SLR2 | SLR3 | SLR1 | SLR2 | SLR3 |
| 0 | **seed set** | **4.43 (22/497)** | **1.71 (16/935)** | **2.29 (11/480)** | **43.14 (22/51)** | **80.00 (16/20)** | **78.57 (11/14)** | **8.03** | **3.35** | **4.45** |
| 1 | forward | 3.97 (34/856) | 1.52 (17/1116) | 2.18 (12/551) | 66.67 (34/51) | 85.00 (17/20) | 85.71 (12/14) | 7.50 | 2.99 | 4.25 |
| | backward | 4.28 (38/887) | 1.46 (20/1374) | 1.99 (14/703) | 74.51 (38/51) | 100.00 (20/20) | 100.00 (14/14) | 8.10 | 2.87 | 3.91 |
| | **union** | **3.72 (46/1238)** | **1.29 (20/1555)** | **1.81 (14/773)** | **90.20 (46/51)** | **100.00 (20/20)** | **100.00 (14/14)** | **7.14** | **2.54** | **3.56** |
| 2 | forward | 3.27 (49/1497) | 1.27 (20/1569) | 1.57 (14/894) | 96.08 (49/51) | 100.00 (20/20) | 100.00 (14/14) | 6.33 | 2.52 | 3.08 |
| | backward | 3.05 (47/1541) | 1.27 (20/1576) | 1.73 (14/811) | 92.16 (47/51) | 100.00 (20/20) | 100.00 (14/14) | 5.90 | 2.51 | 3.39 |
| | **union** | **2.78 (50/1796)** | **1.26 (20/1590)** | **1.50 (14/932)** | **98.04 (50/51)** | **100.00 (20/20)** | **100.00 (14/14)** | **5.41** | **2.48** | **2.96** |
| 3 | forward | 2.82 (51/1806) | - | - | 100.00 (51/51) | - | - | 5.49 | - | - |
| | backward | 2.71 (50/1848) | - | - | 98.04 (50/51) | - | - | 5.27 | - | - |
| | **union** | **2.74 (51/1858)** | **-** | **-** | **100.00 (51/51)** | **-** | **-** | **5.34** | **-** | **-** |
| 4 | forward | 2.74 (51/1860) | - | - | 100.00 (51/51) | - | - | 5.34 | - | - |
| | backward | 2.73 (51/1871) | - | - | 100.00 (51/51) | - | - | 5.31 | - | - |
| | **union** | **2.72 (51/1873)** | **-** | **-** | **100.00 (51/51)** | **-** | **-** | **5.30** | **-** | **-** |

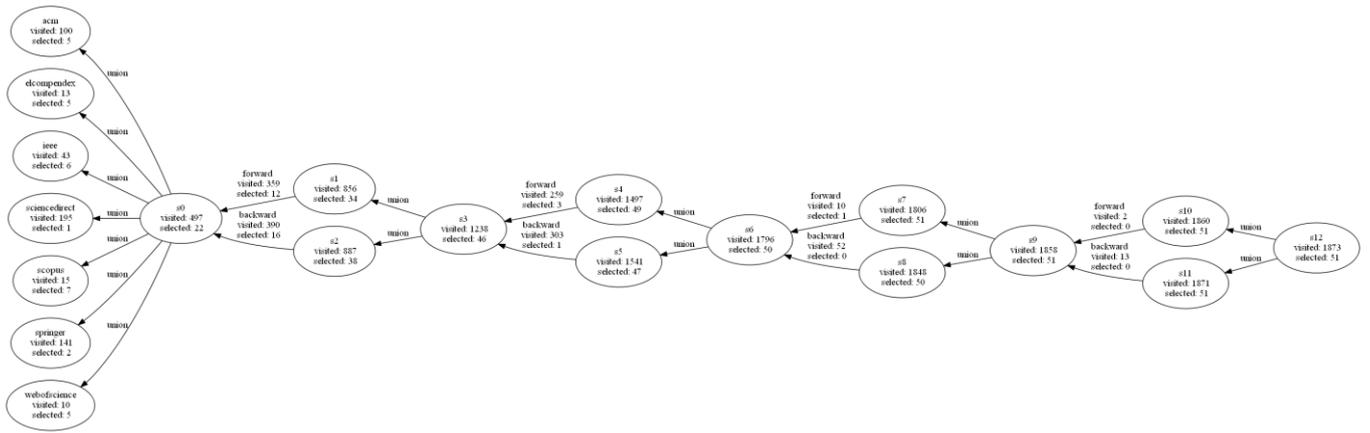

Figure 1 – Overview of the exhaustive database search followed by iterative snowballing for SLR1.

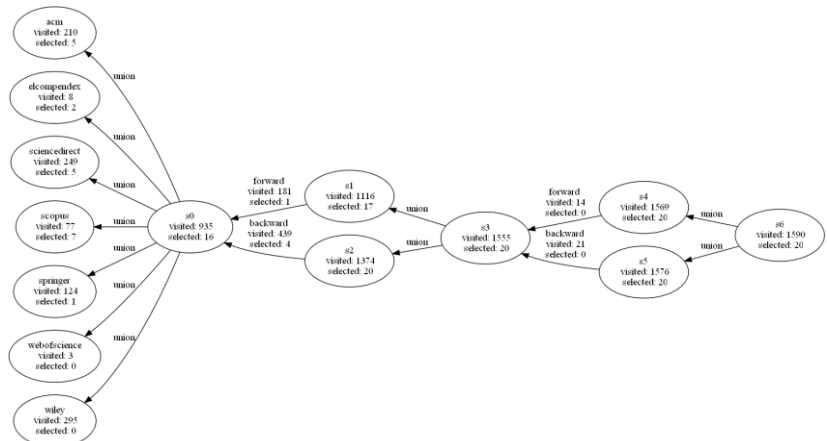

Figure 2 – Overview of the exhaustive database search followed by iterative snowballing for SLR2.

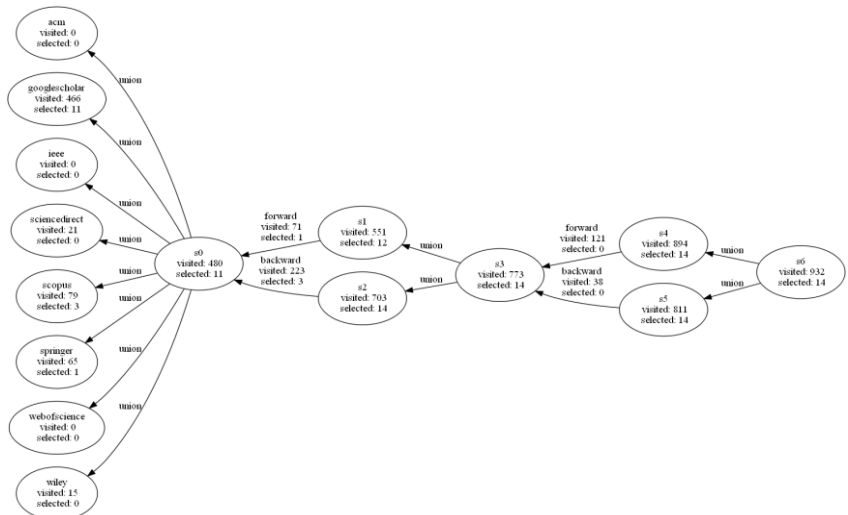

Figure 3 – Overview of the exhaustive database search followed by iterative snowballing for SLR3.

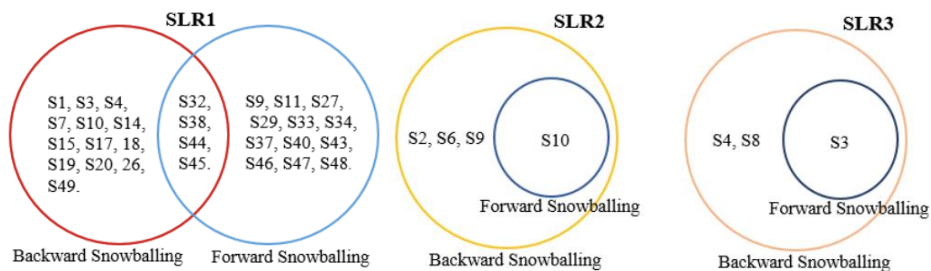

Figure 4 - Venn diagrams contrasting BS and FS.

Moreover, we compute the recall of each strategy based on the total number of articles retrieved by each SLR. Although we cannot guarantee that all possible articles were correctly obtained by each SLR, we assessed the quality of the SLRs to minimize this threat. Additionally, all three SLRs were published in relevant venues, that adopt a serious peer-review process.

Additionally, we are aware that the effort to exclude papers varies depending on how the paper can be excluded. For example, some papers can be easily spotted as irrelevant to the SLR by looking at the paper title or abstract, while others are relevant for the area and might only be excluded after some effort involving a thorough look at the abstract or even the paper content. The number of papers excluded in different steps may vary for the different search strategies, and hence affect the effort for the different strategies. Moreover, some are unrelated noise or duplicates. In this study, we did not consider this relative exclusion effort. Instead, we limited ourselves to calculating precision, recall, and F-measure, which do not consider this type of information. This decision was taken mainly because our selected SLR corpus had no information on how papers were excluded. Fine-tuning this study by modeling and assessing such relative effort represents an opportunity for future work.

With respect to **internal validity**, papers that describe SLRs are usually superficial, not having enough information to allow a precise reproduction of the results. For example, some do not list duplicates or indicate the digital library that returned each article. To mitigate this threat, we contacted the authors and requested the complete review package containing the list of articles returned by each digital library. The paper describing SLR1 indicates that 517 articles were obtained through the database search, and 495 articles remained after duplicate removal. However, we identified, in the spreadsheet provided by the authors, two articles that were not duplicates. Therefore, the total count of articles after removing the duplicates was 497. In SLR2, we also found one non-duplicate paper in a spreadsheet provided by the authors. Thus, the total count of articles was 935 instead of 934. Finally, SLR3 listed a total of 571 papers, including duplicates, but two of them were not present in the spreadsheet. After removing duplicates, the total count of articles was 480.

Moreover, SLR3 did not list duplicates for all digital libraries. It just indicated the first digital library that returned the articles. To mitigate this problem, we reproduced the original search in all other digital libraries to identify which of them also would have returned duplicates of the articles.

When running the *SB Search* strategy, we followed the guidelines provided by Wohlin [2]. However, the guidelines have a subjective step of informal Google Scholar search, which would compromise the reproducibility of our study. Aiming at mitigating this threat, we always considered the top-60 results provided by Google Scholar. Since the SLRs have at most 51 selected articles, this number would be large enough to accommodate all selected articles (recall = 100%), should that be the case.

SLR1 adopted a quality control step at the end of the process, removing 18 studies and three not available articles from the already 51 selected articles. Although the final count reported in their paper is 30 articles, we decided to consider all 51 articles in our analyses. Working with the smaller set would incorrectly affect the precision and recall, as many of the visited articles in the snowballing were there because of the removed articles in the quality control step. Similarly, the SLR2 also adopted a quality control step at the end, removing 12 articles from the 20 already selected articles. We again considered all the 20 selected articles to compute precision and recall, for the same reasons as for SLR1.

Regarding **conclusion validity**, we did not adopt statistical tests during data analysis due to the size of our sample (three SLRs). Consequently, although our results allow observing the performance of the different search strategies on the selected SLRs, they are not conclusive.

Concerning **external validity**, we searched for published SLRs in the SE area that have used database search and snowballing to compose the corpus. Due to the small sample, consisting of only three SLRs in the field of software processes, our results may not be generalizable to all other fields of software engineering. We suggest as future work additional replications of this study over a greater number of SLRs from other areas within software engineering.

Table 9 - Performance of the hybrid strategies.

| Strategy | Precision % | | | Recall % | | | F-measure % | | |
| --- | --- | --- | --- | --- | --- | --- | --- | --- | --- |
| | SLR1 | SLR2 | SLR3 | SLR1 | SLR2 | SLR3 | SLR1 | SLR2 | SLR3 |
| **DB Search** | 4.43 (22/497) | 1.71 (16/935) | 2.29 (11/480) | 43.14 (22/51) | 80.00 (16/20) | 78.57 (11/14) | 8.03 | 3.35 | 4.45 |
| **SB Search** | 3.35 (36/1076) | 1.26 (6/478) | 2.25 (11/489) | 70.59 (36/51) | 30.00 (6/20) | 78.57 (11/14) | 6.39 | 2.41 | 4.37 |
| **DB Search + BS*FS** | 2.72 (51/1873) | 1.26 (20/1590) | 1.50 (14/932) | 100.00 (51/51) | 100.00 (20/20) | 100.00 (14/14) | 5.3 | 2.48 | 2.96 |
| **Scopus + BS*FS** | 3.75 (44/1174) | 1.89 (11/581) | 2.19 (11/502) | 86.27 (44/51) | 55.00 (11/20) | 78.57 (11/14) | 7.18 | 3.66 | 4.26 |
| **Scopus + BS\|\|FS** | 6.51 (19/292) | 2.65 (10/378) | 3.72 (9/242) | 37.25 (19/51) | 50.00 (10/20) | 64.29 (9/14) | 11.08 | 5.03 | 7.03 |
| **Scopus + BS+FS** | 6.19 (35/565) | 2.65 (10/378) | 2.59 (11/424) | 68.63 (35/51) | 50.00 (10/20) | 78.57 (11/14) | 11.36 | 5.03 | 5.02 |
| **Scopus + FS+BS** | 5.81 (24/413) | 1.89 (11/581) | 3.27 (9/275) | 47.06 (24/51) | 55.00 (11/20) | 64.29 (9/14) | 10.34 | 3.66 | 6.23 |

## 8. Related Work

Many other studies contrast different SLR search strategies. In this section, we present some of this related work and contrast it with our approach.

Jalali and Wohlin [4] conducted a study to compare two different search approaches: the use of database search and the use of snowballing in the same SLR. They observed similar results for both SLR search strategies.

Wohlin [2] proposes guidelines for snowballing and assesses the guidelines through the replication of a published SLR that used only database search. As a conclusion, snowballing may be a potential alternative to database searches.

Badampudi *et al.* [5] applied snowballing in a study, evaluated the efficiency and reliability of snowballing, and compared it with a database search strategy. They concluded that the efficiency of snowballing is comparable to the efficiency of database searches.

Wohlin [20] compared snowballing with a database search update (i.e., two similar database search SLRs covering different periods). They concluded that both approaches are comparable when it comes to the papers found, although snowballing is more efficient.

Felizardo et al. [21] compare outcomes of an SLR update using forward snowballing versus database search. Although database search reached higher recall, forward snowballing reached significantly higher precision. Consequently, forward snowballing has the potential to reduce the effort in updating SLRs in SE. Mendes et al. [22], also evaluate the use of different search strategies for updating SLRs and provide specific recommendations for the SLR update context.

Felizardo et al. [23] evaluate the use of different databases for applying forward snowballing to update SLRs. They concluded that the use of a specific database is not recommended for forward snowballing in order to update an SLR, since relevant studies may not be found. However, using a generic (Scopus or Google Scholar) database is sufficient to find the studies.

Kitchenham et al. [24] compared the use of manual search with broad automated database searches. They found that broad automated searches were able to find more studies than manual searches, but eventually with poor quality.

Skoglund and Runeson [25] presented a reference-based search strategy that checks whether papers cite other papers together. They evaluated their strategy over three published SLRs and observed significant variation in the results.

Dieste and Padua [26] analyzed the effects of adding a few or many terms to queries on the sensitivity and precision of the SLR. They concluded that optimizing search strategies is not a straightforward task.

MacDonell *et al* [28] investigated how consistent the process adopted in SLRs are, and the effects on the stability of outcomes. They compared the results of two independent SLRs. They concluded that groups of researchers with similar domain experience could reach the same outcomes.

Differently from the aforementioned studies, which in most cases directly contrasted database search with snowballing, our study went further and focused on specific aspects of database search and snowballing. Regarding database search, we investigated the performance (both actual and potential) of different digital libraries and how they complement each other. Concerning snowballing, we investigated whether multiple iterations are necessary and how forward snowballing compares to backward. snowballing Finally, we also proposed four hybrid search strategies and contrasted their performance.

## 9. Conclusions and future work

In this work, we proposed and evaluated hybrid search strategies that combine database searches with snowballing. We could observe that Scopus is the most consistent option in achieving high recall, but it found just from 13% to 35% of the relevant papers alone. Complementing Scopus with ACM Digital Library is an appropriate choice. Scopus and ACM Digital Library together found from 23% to 60% of the relevant papers. Moreover, investing extra effort on the elaboration of the search string is also worth it – the recall of Scopus could have raised from the range of 13% to 35% to the range of 50% to 95% depending on the search string.

Nevertheless, we could also observe that a single iteration of backward and forward snowballing complementing the database search provides 90% to 100% of recall. When choosing between backward and forward snowballing, one should keep in mind that forward snowballing is a prominent option to improve precision and backward snowballing is a prominent option to improve recall.

All in all, *Scopus + BS||FS*, *Scopus + BS+FS*, and *Scopus + FS+BS* are efficient hybrid search strategies in comparison to plain *DB Search* strategy and *DB Search + BS\*FS*. Depending on the goals of the SLR and the resources available, one of these hybrid strategies may be an appropriate alternative.

As a conclusion, the use of hybrid search strategies may contribute to SLR effort reduction as they reduce the number of papers that are not relevant to the goals of the SLR. However, in our study we did not measure time and have no direct evidence that looking at fewer papers, in fact, reduces effort. Indeed, some papers are easily discarded based on title in a reference list, while others demand a complete read before being discarded.

As future work, due to this study's small sample, consisting of only three SLRs in the field of software processes, we envision the replication of this study over additional SLRs. We also suggest the inclusion of other search strategies in future analyses, such as manual search, author-based search, and venue-based search. Finally, there is a need to also take effort into account in the comparisons of search strategies.

**Acknowledgments**

We thank the Brazilian Research Council (CNPq) for financial support.